\RequirePackage{rotating}
\documentclass[useAMS,usenatbib]{mn2e}

\usepackage{graphics}
\usepackage{graphicx}
\usepackage{amsxtra}
\usepackage{journals}
\usepackage{url}
\usepackage{latexsym}
\usepackage{amssymb}
\usepackage{times}
\usepackage{amsmath}
\usepackage{subfigure}
\usepackage{natbib}
\usepackage[colorlinks=true, linkcolor=blue, citecolor=blue, urlcolor=blue]{hyperref}
\usepackage{soul,color}
\usepackage{txfonts}
\usepackage{rotating}
\usepackage{lscape} 
\usepackage{threeparttable} 
\usepackage{mathtools} 
\usepackage{ulem}

\usepackage{color}

\begin{document}
\title
[On the discrepancy between parallaxes]
{On the discrepancy between asteroseismic and Gaia DR1 TGAS parallaxes}
\author[Gontcharov \& Mosenkov.]
{
George~A.~Gontcharov$^{1,2}$\thanks{E-mail: george.gontcharov@tdt.edu.vn} and Aleksandr~V.~Mosenkov$^{3,4,5}$
\newauthor
\\ \\
$^{1}$Department for Management of Science and Technology Development,
Ton Duc Thang University, Ho Chi Minh City, Vietnam\\
$^{2}$Faculty of Applied Sciences, Ton Duc Thang University, Ho Chi Minh City, Vietnam\\
$^{3}$Sterrenkundig Observatorium, Universiteit Gent, Krijgslaan 281, B-9000 Gent, Belgium\\
$^{4}$St.Petersburg State University, 7/9 Universitetskaya nab., St.Petersburg, 199034 Russia\\
$^{5}$Central Astronomical Observatory, Russian Academy of Sciences, 65/1 Pulkovskoye chaussee, St. Petersburg, 196140 Russia\\
}

\date{\today}

\pagerange{\pageref{firstpage}--\pageref{lastpage}} \pubyear{2017}

\maketitle

\label{firstpage}

\begin{abstract}
Recently, a deviation of the Gaia TGAS parallaxes from the asteroseismic ones for giants was found. 
We show that for parallaxes $\varpi<1.5$ mas it can be explained by a selection effect in favour of bright and luminous giants in the Tycho-2 and TGAS catalogues.
Another explanation of this deviation seems to be valid for $\varpi>1.5$ mas based on the best extinction estimates: the deviation may be caused not by a bias of parallax, but by an underestimation of the extinction (and, consequently, an overestimation of the calculated absolute magnitude) in the asteroseismic results.
We demonstrate that the reliable estimates of the reddening and extinction (about 0.22 mag of the visual extinction for the Kepler field) better fit both the giants and main
sequence stars to the PARSEC, MIST and YaPSI isochrones.
\end{abstract}


\section{Introduction}
\label{sec.intro}
Recently, the positions from the Hipparcos \citep{hip2} and Tycho-2 \citep{tycho2} catalogues have been combined with early data of the Gaia mission \citep{gaia3} and presented as
the Gaia DR1 Tycho-Gaia astrometric solution \citep[TGAS, ][]{tgas} with the parallaxes $\varpi$ for two million stars.

The comparisons summarized by \citet{gaia3} have shown the offsets $\Delta\varpi$ in the sense ``TGAS minus other source'' from
$-0.25\pm0.05$ mas for 111 eclipsing binaries \citep{stassun} to $0.00\pm0.05$ mas for about 100 RR Lyrae stars \citep{sesar}.
This contradictory information on a possible TGAS parallax bias needs further testing.
The giants, especially the clump giants, seem to be some of the most suitable stars for such tests
because they are luminous, numerous in TGAS, and accurately positioned in the Hertzsprung-Russell (HR) diagram.

Recently, a discrepancy has been found by \citet{deridder} between the TGAS and asteroseismic parallaxes determined for giants by \citet[][hereafter RGM]{rodrigues}.
\citet{deridder} suggested that either the TGAS parallaxes are biased, or the asteroseismic parallaxes are affected by  inaccurate interstellar extinction corrections
and/or poorly known bulk metallicities, which would introduce systematics into the estimated stellar luminosities.
\citet[][hereafter DLM]{davies} analysed the results of RGM and \citet{deridder} using a subsample of their clump giants observed by the NASA {\it Kepler} 
mission \citep{2004SPIE.5487.1491K}.
DLM found a median offset of the TGAS parallaxes of $-0.1$ mas (which leads to an underestimation of the parallaxes and overestimation of the distances).
However, only few clump giants have $\varpi>1.5$ and they show little if any offset.

We point out a discrepancy between their extinction estimates.
DLM obtained the mean extinction $\overline{A_\mathrm{K_s}}\approx0.02^m$ for the {\it Kepler} field from the 3D extinction map of \citet[][hereafter GSF]{green}. 
Using the extinction law of \citet{cardelli}, this translates into $\overline{A_\mathrm{V}}\approx0.17^m$, whereas RGM obtained $\overline{A_\mathrm{V}}<0.11^m$.

The mentioned above studies suggested that the red clump giants are robustly separated by asteroseismology from the branch, asymptotic branch and secondary clump giants.
These types differ by the nuclear fusion inside the stars:
core helium for the clump and secondary clump,
envelope hydrogen for the branch, and
both constituents for the asymptotic branch.
The majority of the clump giants (low-mass old stars) have a degenerated core before the helium fusion.
The minority (higher-mass young stars, the secondary clump) are massive enough to have ignited helium in a non-degenerate core.
Therefore, the former has quite a narrow range of the absolute magnitude which is almost independent of colour, age, and metallicity for the near-infrared bands.
The latter has a much wider range of the absolute magnitude \citep{rcg}.

Let us consider two sets of asteroseismic distances and extinctions ($R_\mathrm{direct}$, $A_\mathrm{V direct}$ and $R_\mathrm{Bayes}$, $A_\mathrm{V Bayes}$) calculated by RGM 
using, respectively, the direct method (assuming no knowledge of the evolutionary state of the star), and a model-dependent Bayesian approach (by forcing the estimated luminosities to some assumed values).
The asteroseismically derived luminosity and effective temperature were fitted via spectral energy distribution (SED) to the photometry from 
the Sloan Digital Sky Survey \citep[SDSS,][]{Eisenstein2011}, 
the Two Micron All Sky Survey \citep[2MASS,][]{2mass} and the Wide-field Infrared Survey Explorer \citep[WISE,][]{wise} catalogues.

Also, we consider two sets of distances based on the TGAS parallaxes. $R_\mathrm{TGAS}$ are calculated as $1/\varpi$ and
$R_\mathrm{modemw}$ were found by \citet{bailer3} using $\varpi$, together with their errors $\sigma(\varpi)$.

\begin{figure*}
\includegraphics{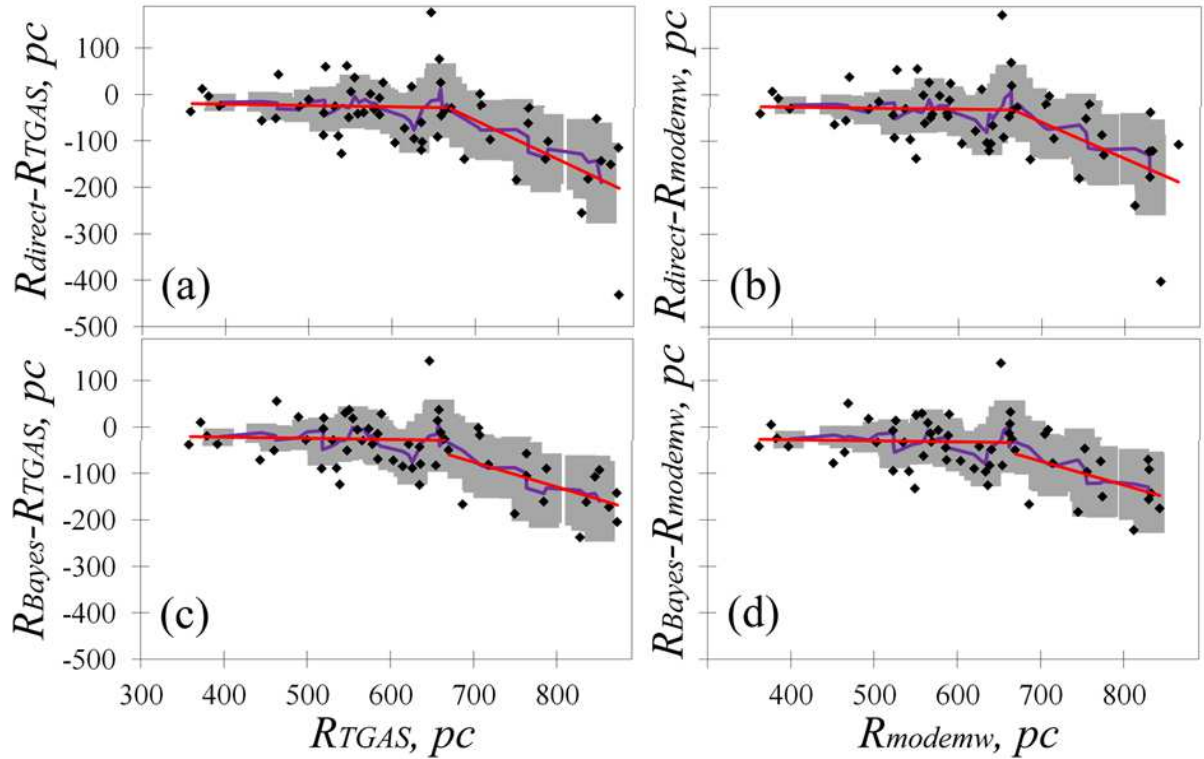}
\caption{The differences between $R_\mathrm{direct}$, $R_\mathrm{Bayes}$ and $R_\mathrm{TGAS}$, $R_\mathrm{modemw}$ 
in dependence on $R_\mathrm{TGAS}$ and $R_\mathrm{modemw}$
for common giants with $\sigma(\varpi)/\varpi<0.3$, $B_\mathrm{T}<11.5^m$, and $V_\mathrm{T}<10.5^m$.
The red lines denote the trends inside and outside the distance 670 pc.
The purple curves show the moving average over 7 points, with the standard deviation of the average shown by the grey bars.
}
\label{dist}
\end{figure*}

For the giants discussed in DLM, Fig.~\ref{dist} shows various differences between the asteroseismic ($R_\mathrm{direct}$ and $R_\mathrm{Bayes}$) and 
trigonometric ($R_\mathrm{TGAS}$ and $R_\mathrm{modemw}$) distances, in dependence on trigonometric distances. 
For higher accuracy we select stars with $B_\mathrm{T}<11.5^m$, $V_\mathrm{T}<10.5^m$, and $\sigma(\varpi)/\varpi<0.3$.
The red lines in all plots denote the trends inside and outside the distance 670 pc.
The purple curves represent the moving average over 7 points, with the standard deviation of the average shown by the grey bars.
We can see the same pattern in all 4 plots in Fig.~\ref{dist}.
For $R<670$ pc ($\varpi>1.5$ mas) all the distances exhibit small constant offsets of 
$R_\mathrm{TGAS}-R_\mathrm{direct}=9$, 
$R_\mathrm{modemw}-R_\mathrm{direct}=18$,
$R_\mathrm{TGAS}-R_\mathrm{Bayes}=12$ pc,
and
$R_\mathrm{modemw}-R_\mathrm{Bayes}=19$ pc.
It is well seen that for $R_\mathrm{TGAS}>670$ pc the absolute values of all these differences increase with distance.

The reason for this increase is obvious. The TGAS dataset is based on Tycho-2, and, therefore, it is complete for magnitudes $V_\mathrm{T}\lesssim11.5^m$.
Taking into account an approximate range of the absolute magnitude for the clump as $-0.5^m<M_\mathrm{V_T}<1.8^m$ and an extinction estimate as $A_\mathrm{V_T}<0.5$,
one can see the completeness of the TGAS subsample of the clump stars only up to about 690 pc \citep{g2016}.
This has been proven by \citet{rcg} who selected and analysed an almost complete sample of the 97348 clump giants from the Tycho-2 catalogue.
Therefore, the trends seen in Fig.~\ref{dist} for distances $R_\mathrm{TGAS}>670$ pc are the direct result of an increasing incompleteness of the sample. One has to consider only a brighter part of the clump. Assuming its average absolute magnitude as the one for the whole clump, obviously, an overestimated average distance will be retrieved.
Thus, an important conclusion should be drawn: 
to reveal a TGAS parallax bias, only complete samples of the clump giants with $\varpi>1.5$ mas are suitable. 
Evidently, ``a clear systematic error in the TGAS parallaxes versus distance'', which was proposed by DLM, is a result of this selection effect.
Almost all their clump giants have $\varpi<1.6$ and, naturally, overestimated distances.

Further in this study we analyze the data for the giants with the estimated asteroseismic and TGAS distances in order to explain their offsets for $\varpi>1.5$ mas.

\section{Extinction}
\label{sec.extinction}

Any discrepancy between the magnitude $m$, absolute magnitude $M$, extinction $A$, and distance $R$ in the equation
\begin{equation}
\label{basic}
m=M+A-5+5\log(R)\, ,
\end{equation}
may be due to wrong estimates of $M$, $A$, or $R$.

Since in the range $R<670$ pc the sample contains few clump giants and many giants of other types, we pay more attention to $R_\mathrm{direct}$ and $A_\mathrm{V direct}$
which were obtained by the direct method, assuming no knowledge of the evolutionary state of the star.
Hereafter, we show only results with $R_\mathrm{direct}$ (together with $A_\mathrm{V direct}$) and $R_\mathrm{TGAS}$.
It is evident from Fig.~\ref{dist} (and has been verified by us) that the use of $R_\mathrm{Bayes}$ (together with $A_\mathrm{V Bayes}$) and $R_\mathrm{modemw}$ 
provides similar results.

\begin{figure}
\includegraphics{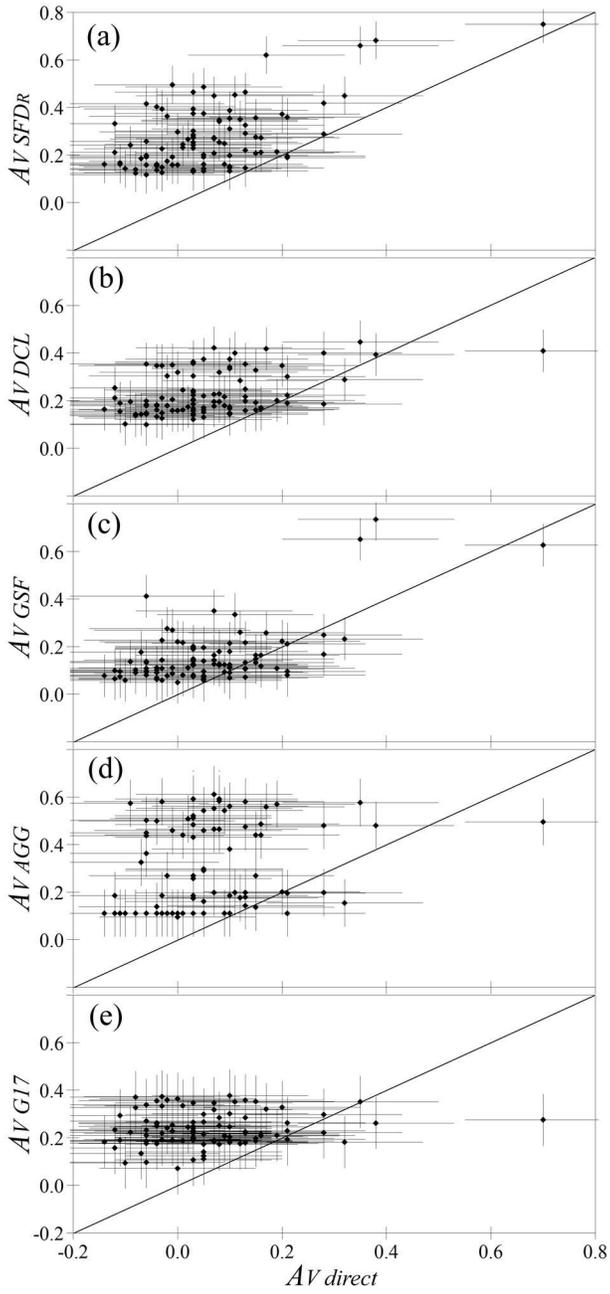}
\caption{Extinction $A_\mathrm{V direct}$ for the giants in the {\it Kepler} field with $\varpi>1.5$ mas versus the extinction estimates from SFD$_R$, DCL, GSF, AGG, and G17.
}
\label{av}
\end{figure}

RGM noticed that $A_\mathrm{V direct}$ is considerably lower, on average, than some other estimates of the extinction obtained for the {\it Kepler} field.
Fig.~\ref{av} shows the relation between the $A_\mathrm{V direct}$ and $A_\mathrm{V}$ estimates which are taken from various sources of extinction for 97 giants selected from RGM with $\varpi>1.5$ mas.
The sources are:
a) \citet[][hereafter SFD]{sfd} reduced as described below,
b) \citet[][hereafter DCL]{drimmel},
c)  GSF, 
d) \citet[][hereafter AGG]{arenou}, and
e) \citet[][hereafter G17]{g17}. A brief review of these maps is given in \citet{astroph}.

The SFD map is a 2D emission-based map of the cumulative extinction to infinity.
We reduced it to the distance $R$ following the barometric law \citep[][ p. 265]{parenago}:
\begin{equation}
\label{baro}
A_\mathrm{V R}=A_\mathrm{V}\,(1-\mathrm{e}^{-|Z-Z_0|/Z_\mathrm{A}})\,,
\end{equation}
where $A_\mathrm{V R}$ is the extinction to the distance $R$,
$A_\mathrm{V}$ is the extinction to infinity for the same line of sight,
$Z=R\sin(b)$ is a Galactic coordinate in kpc,
$Z_0$ is the vertical offset of the midplane of the dust layer with respect to the Sun in kpc,
$Z_\mathrm{A}$ is the scale height of the dust layer in kpc.
We accept $Z_0=-13$~pc and $Z_\mathrm{A}=100$~pc. 
The reduced map is hereafter designated as SFD$_R$.
We compared SFD with another 2D emission-based extinction map of \citet{2015ApJ...798...88M}, which is based on the {\it Planck} data \citep{planck2014}, and found that both maps
provide very similar extinction values for the selected giants.

In this work we use the extinction law which was applied in the PARSEC database \citep[][http://stev.oapd.inaf.it/cgi-bin/cmd]{bressan}.  
However, the extinction laws of \citet{cardelli} and \citet{wd2001} provide similar results.
The reddening from G17 is converted into the extinction $A_\mathrm{V}=R_\mathrm{V}\,E(B-V)$ using the 3D map of the spatial variations of the coefficient $R_\mathrm{V}$ \citep{rv}.

Fig.~\ref{av} shows that, as supposed by RGM, all sources provide the extinction estimates which occur much higher than $A_\mathrm{V direct}$. 
The median estimates of the extinction and their errors declared by the corresponding authors for the 97 giants with $\varpi>1.5$ mas in the {\it Kepler} field are:
$A_\mathrm{V direct}=0.03^m\pm0.15^m$, $A_\mathrm{V Bayes}=0.07^m\pm0.08^m$, 
$A_\mathrm{V SFD_R}=0.24^m\pm0.08^m$, $A_\mathrm{V DCL}=0.20^m\pm0.09^m$, $A_\mathrm{V GSF}=0.12^m\pm0.09^m$, $A_\mathrm{V AGG}=0.20^m\pm0.10^m$,
$A_\mathrm{V G17}=0.23^m\pm0.12^m$.
Thus, these estimates can be separated into two incompatible groups: lower estimates $A_\mathrm{V direct}$, $A_\mathrm{V Bayes}$, and $A_\mathrm{V GSF}$ versus higher estimates 
$A_\mathrm{V SFD_R}$, $A_\mathrm{V DCL}$, $A_\mathrm{V AGG}$, and $A_\mathrm{V G17}$.

In order to choose proper estimates of the extinction for the selected giants, we put them on the HR diagram with respect to the PARSEC,
MIST \citep[][http://waps.cfa.harvard.edu/MIST/]{mist} and YaPSI \citep[][http://www.astro.yale.edu/yapsi/]{yapsi} theoretical isochrones.
However, this positioning is not very sensitive to the reddening and extinction in the diagram's domain of the giants.
We, therefore, add to the diagram the early-type main sequence (ETMS) stars whose positioning is more sensitive to the reddening and extinction.
In order to consider the ETMS stars along with the giants at exactly the same distances, longitudes, and latitudes, we choose for the diagram 
the absolute magnitude $M_\mathrm{V_T}$ which is similar for the giants and ETMS stars of the classes B and A. We consider $M_\mathrm{V_T}$ versus 
$(B_\mathrm{T}-V_\mathrm{T})_0$ colour based on the Tycho-2 photometry, taking into account, as mentioned above,
that the Tycho-2 and TGAS catalogues contain complete samples of these stars to $R\approx670$ pc.

\begin{figure*}
\includegraphics{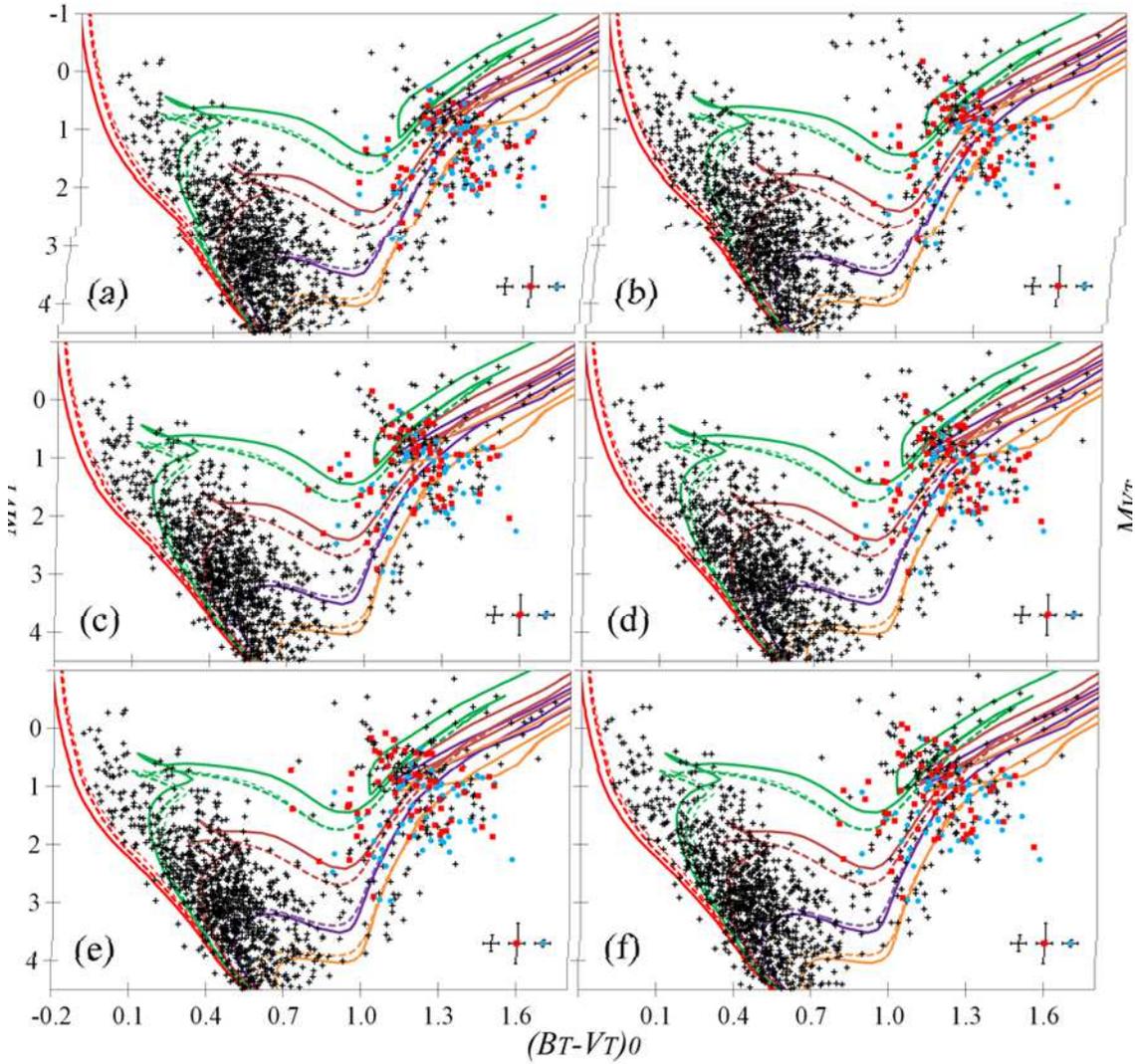}
\caption{The HR diagram $(B_\mathrm{T}-V_\mathrm{T})_0 - M_\mathrm{V_T}$ based on $R_\mathrm{TGAS}$ for the 1225 TGAS stars in the {\it Kepler} field 
with $\sigma(\varpi)/\varpi<0.1$ -- black crosses and the
56 giants from RGM with $\varpi>1.5$ mas, $\sigma(\varpi)/\varpi<0.3$, $B_\mathrm{T}<11.5^m$, and $V_\mathrm{T}<10.5^m$ -- red squares
with 
a) zero extinction, 
or the estimates of extinction from 
b) SFD$_R$,
c) DCL,
d) GSF,
e) AGG, and
f) G17.
Positions of the same 56 giants based on $R_\mathrm{direct}$ and $A_\mathrm{V direct}$ estimates -- blue circles.
The PARSEC and MIST isochrones for 0.1, 1, 2, 5 and 10 Gyr (with details described in the text) -- red, green, brown, purple and orange curves, respectively.
}
\label{hr}
\end{figure*}

The resulting HR diagrams are shown in Fig.~\ref{hr}.
The PARSEC isochrones are shown as the solid lines of different colours:
100 Myr, $\mathbf Z=0.0152$, $\mathbf Y=0.2756$ -- red,
1 Gyr, $\mathbf Z=0.0152$, $\mathbf Y=0.2756$ -- green,
2 Gyr, $\mathbf Z=0.0152$, $\mathbf Y=0.2756$ -- brown,
5 Gyr, $\mathbf Z=0.012$, $\mathbf Y=0.27$ -- purple and
10 Gyr, $\mathbf Z=0.012$, $\mathbf Y=0.27$ -- orange.
The MIST isochrones drawn to the tip of the giant branch are the thin and thick dashed lines for the cases with and without rotation of the star, respectively:
100 Myr, initial $\mathbf Z=0.0142$, $\mathbf Y=0.2703$, actual $\mathbf Z\approx0.0156$ -- red,
1 Gyr, initial $\mathbf Z=0.0142$, $\mathbf Y=0.2703$, actual $\mathbf Z\approx0.0156$ -- green,
2 Gyr, initial $\mathbf Z=0.0142$, $\mathbf Y=0.2703$, actual $\mathbf Z\approx0.0156$ -- brown,
5 Gyr, initial $\mathbf Z=0.01$, $\mathbf Y=0.2653$, actual $\mathbf Z\approx0.012$ -- purple and
10 Gyr, initial $\mathbf Z=0.01$, $\mathbf Y=0.2653$, actual $\mathbf Z\approx0.012$ -- orange
(metallicity changes from initial to actual as the star evolves).
The rotation is important only for massive stars.
For the isochrones we adopt the average age--metallicity relation of \citet{haywood}.
The blue circles are 56 giants from RGM with $\varpi>1.5$ mas, $\sigma(\varpi)/\varpi<0.3$, $B_\mathrm{T}<11.5^m$, and $V_\mathrm{T}<10.5^m$ with the
$R_\mathrm{direct}$ and $A_\mathrm{V direct}$ estimates.
The red squares are the same 56 giants but positioned by use of $R_\mathrm{TGAS}$ and the estimates of $A_\mathrm{V_T}$ and $E(B_\mathrm{T}-V_\mathrm{T})$ which are taken
from the extinction sources:
a) zero reddening and extinction, 
b) SFD$_R$,
c) DCL,
d) GSF,
e) AGG, and
f) G17.
The black crosses are all 1225 TGAS stars with $\sigma(\varpi)/\varpi<0.1$ in the {\it Kepler} field ($68^{\circ}<l<85^{\circ}$, $5^{\circ}<b<22^{\circ}$)
positioned by use of $R_\mathrm{TGAS}$ and the estimates of $A_\mathrm{V_T}$ and $E(B_\mathrm{T}-V_\mathrm{T})$ from the same extinction sources.
The three crosses in the right bottom corner in each plot show the error bars of the positioning due to the uncertainties of the distance and the photometry.

\begin{figure}
\includegraphics{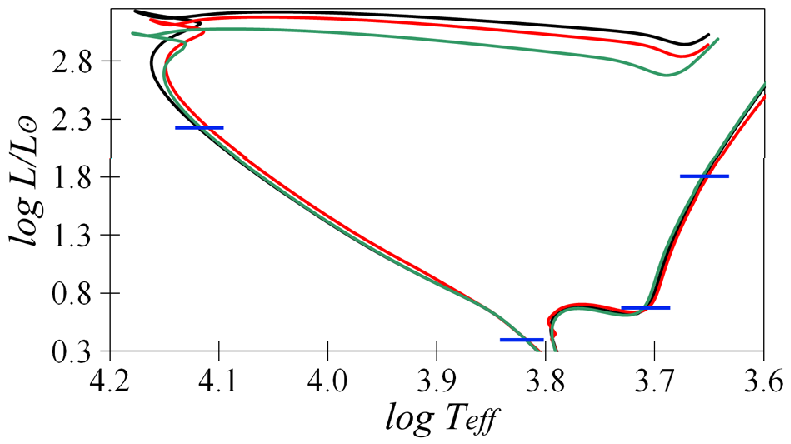}
\caption{The PARSEC, MIST and YaPSI isochrones (see text) in the HR diagram $\log T_{eff} - \log L/L_{\odot}$ -- black, red and green lines, respectively.
The blue lines show the luminosity ranges described in the text.}
\label{tl}
\end{figure}

Unfortunately, YaPSI have not provided $B_\mathrm{T}$ and $V_\mathrm{T}$ isochrones. Nevertheless, in Fig.~\ref{tl} we show the YaPSI isochrones for 100 Myr, $\mathbf Z=0.0162$, $\mathbf Y=0.28$ and for
5 Gyr, $\mathbf Z=0.013$, $\mathbf Y=0.28$ by the green curves in the HR diagram ``effective temperature -- luminosity'', together with the PARSEC (black) and MIST (red)
isochrones of the same ages indicated in Fig.~\ref{hr} (with stellar rotation for MIST).
The blue lines define the most interesting luminosity ranges which correspond to $0^m<M_\mathrm{V_T}<4^m$ for the ETMS and $1^m<M_\mathrm{V_T}<3.5^m$ for the giant branch.
For the branch stars the isochrones almost coincide, whereas for the ETMS the MIST isochrones deviate from the others both in Fig.~\ref{hr} and \ref{tl} up to 
$\Delta((B_\mathrm{T}-V_\mathrm{T})_0)=0.06^m$.
A considerable part of this discrepancy is due to the rotation of stars taken into account only in the MIST database.

Fig.~\ref{hr} shows that RGM put the majority of their giants (blue circles) at the part of the branch fainter than the clump, along the 10 Gyr isochrones.
It demonstrates a little (if any) young secondary clump and the asymptotic branch giants (along the green line), as well as the branch giants brighter than the clump.
The positions of the blue circles in different plots show that $A_\mathrm{V direct}$ produces the HR diagram which is indistinguishable from the zero-extinction one.
However, a non-zero extinction is evident, as:

Firstly, too few ETMS stars near 100 Myr and to the left of 1 Gyr isochrones in the plot (a) contradict the Besan\c{c}on Galaxy model \citep[][hereafter BGM]{bmg2}.
A noticeable lack of the ETMS stars is seen also for the GSF map.

Secondly, the lower part of the giant branch (lower than the clump) should look approximately symmetric around the 5 Gyr isochrone
in case of the constant star formation rate (SFR) in the Galactic disc within the last 10 Gyr.
For variable SFR we can test the ratio of the numbers of the branch giants on the both sides of the 5 Gyr isochrone.
This ratio should be $1.5-2.5$ for the decreasing SFRs of \citet{aumer} and \citet{jj} which have been considered for the BGM.
The actual ratios are 3.4, 2.0, 2.5, 2.8, 2.3, 2.0 for the zero extinction, SFD$_R$, DCL, GSF, AGG and G17 estimates, respectively, as shown for all 1225 TGAS stars
(the 56 giants from RGM may not be suitable for this test due to the selection).
Thus, the 5 Gyr lower branch does not fit the reliable SFR in cases of the zero extinction and GSF.
It should be noticed that all the derived ratios are far from 1. It favours a decreasing SFR in agreement with the BGM model.

Both reasons allow us to conclude that for the {\it Kepler} field the lowest extinction estimates from RGM and GSF are worse
than the highest extinction estimates from SFD$_R$, DCL, AGG, and G17.
Their average $A_\mathrm{V}=0.22^m\pm0.02^m$ is the most probable estimate.
For the most reliable estimates, Fig.~\ref{hr} demonstrates a considerable change of the evolutionary state of the giants (the red squares) with respect to the one 
proposed by RGM (the blue circles).
Now the majority of these giants are distributed more uniformly among all ages, with a considerable number of stars younger than 2 Gyr.
Such a large fraction of young and massive giants may be explained by the fact that the space under consideration belongs to the Local Orion spiral arm.

\section{Conclusions}
\label{conclusions}

For the giants, the deviation of the TGAS parallaxes $\varpi<1.5$ mas from the asteroseismic ones may be explained by the selection effect in favor of bright and, consequently, luminous giants in the Tycho-2 and TGAS catalogues.
For the space within $\varpi>1.5$ mas we have considered five sources of extinction estimates to put the giants and main sequence stars among the PARSEC, MIST and YaPSI isochrones.
This shows that the reason of the discrepancy between the TGAS and asteroseismic data for $\varpi>1.5$ mas seems to be not a bias of parallax, 
but an underestimation of the extinction in the asteroseismic results and, consequently, an overestimation of the absolute magnitude.

\section{Acknowledgements}

AM is partly supported by the Russian Foundation for Basic Researches (grant number 14-02-810).
AM is a beneficiary of a mobility grant from the Belgian Federal Science Policy Office.
We thank Justin Turman for the improvement of the style of the paper.

\bibliographystyle{mn2e}
\bibliography{bias1}

\label{lastpage}
\end{document}